\documentstyle[aps,preprint,version2]{revtex}
\begin{document}
\draft

\begin{title}
Electronic and structural properties of GaN by the \\
full-potential LMTO method : the role of the $d$ electrons
\end{title}

\author{Vincenzo Fiorentini, Michael Methfessel, and Matthias Scheffler}
\begin{instit}
Fritz-Haber-Institut der Max-Planck-Gesellschaft,
Faradayweg 4-6, D-1000 BERLIN 33, Germany
\end{instit}

\receipt{26 August 1992}

\begin{abstract}
The structural and electronic properties of cubic GaN are studied within
the local density approximation by the full-potential linear muffin-tin
orbitals method. The Ga $3d$ electrons are treated as band states, and no
shape approximation is made to the potential and charge density.
The influence of $d$ electrons on the band structure, charge
density, and bonding properties is analyzed. It is found that due
to the energy resonance of the Ga 3$d$ states with nitrogen 2$s$ states,
the cation $d$ bands are not inert, and features unusual for a III-V compound
are found in the lower part of the valence band
and in the valence charge density and density of states. To clarify
the influence of the Ga $d$ states on the cohesive properties,
additional full and frozen--overlapped-core calculations were performed
for GaN, cubic ZnS, GaAs, and
Si. The results show, in addition to the known importance of non-linear
core-valence exchange-correlation corrections, that an explicit description of
closed-shell repulsion effects is necessary to obtain accurate results for GaN
and similar systems.
In summary, GaN appears to be somewhat exceptional among the
III-V compounds and reminiscent of II-VI materials, in that
its band structure and cohesive properties are sensitive to a proper
treatment of the cation $d$ bands, as a result of the presence of the latter
in the valence band range.
\end{abstract}


\parindent=10pt

\section{INTRODUCTION}

Recently, considerable interest has arisen \cite{rev} in the wide-gap
III-V nitrides as candidates for the realization of light emitting devices in
the blue range of the visible spectrum. In particular, succesful epitaxial
growth of thin films of gallium nitride has recently been demonstrated,
resulting in materials having the wurtzite or the zincblende
structure \cite{expg}, depending on the substrate on which the
growth was effected.

Despite its increasing technological interest, GaN is poorly
known from the theoretical point of view. Accurate theoretical predictions are
of some relevance in this case, since characterization and experimental
work on this material is still at an early stage when compared to the average
developement level of III-V technology. The issue of a theoretical description
of this system
is also of interest in itself, due to the open question about the
role of the Ga $d$ electrons.
In this work we present calculations of lattice properties and band structure
of GaN in the zincblende structure (ZB)
 performed {\it ab initio} within the local-density approximation
(LDA) to density functional theory \cite{gd}\, using the all-electron
full-potential linear muffin-tin orbitals method (FP-LMTO) \cite{fp}.
We analyze the character of band states, charge densities and density of
states, showing the relevance of the Ga $d$ shell in determining the
 properties of the
material.
To demonstrate the effects of the $d$ shell on cohesive properties,
we perform full and frozen--overlapped-core calculations
for GaN, ZnS, GaAs, and Si monitoring the role of the $d$ electrons
and of core relaxation in the four materials.

In Sec.II we discuss technicalities, in Sec.III some background, in Sec.IV
the results are presented and discussed, and in Sec.V we examine
the effects of $d$ electrons on cohesive properties, charge density and
density of states (DOS). Rydberg and
bohr atomic units are used throughout the paper unless otherwise stated.

\section{TECHNICALITIES}

Our calculations were performed within the LDA by the FP-LMTO method
\cite{fp}\,
using the Vosko-Wilk-Nusair \cite{vwn}\, parametrization of the Ceperley-Alder
\cite{ca}\, exchange correlation energy, and
the Monkhorst-Pack 10 special-points mesh \cite{mpa}. The $2s$ and $2p$ states
of N and the $3d, 4s$, and $4p$ states of Ga are treated as bands, whereas the
remaining core states are
self-consistently relaxed in a spherical approximation. As variational basis
for solving the Schr\"odinger equation, three augmented Hankel functions are
used on each occupied atomic site with decay energies --0.7, --1.0 and --2.3
Ryd up to angular momentum $l = 2$.
This LMTO basis is centered only on the Ga and N spheres,
giving 27 functions per atom.

The atomic spheres are non-overlapping (sphere radii are 97 \% of half the
nearest-neighbor distance) in
contrast with the often-used atomic-spheres approximation (ASA) version
of the LMTO method \cite{asa}.
 As usual, empty spheres
(of the same size as the atomic ones) are inserted in the interstitial regions
of the ZB structure to improve the packing fraction.

A full-potential technique needs an accurate evaluation of interstitial
three-center integrals and charge density. This is done by an
interpolation technique which represents the product of two Hankel functions
as a linear combination of other Hankel functions in the interstitial region.
The same technique is used to represent the interstitial exchange-correlation
potential and energy density. More details are to be found in Ref.\cite{fp}.
The auxiliary interstitial charge density is expressed as an expansion in
Hankel functions with decay energies --1.0 and --3.0 Ryd, centered in all
spheres, with angular momentum  up to $l = 4$.

The result presented below were obtained using a non-relativistic code.
We also performed a scalar-relativistic calculation for GaN with the same
method,  and the results are
found to be essentially unchanged (the lattice constant and fundamental gap
are reduced by 0.2 \% and 1.5 \%  respectively, and the cohesive energy is
practically identical).

\section{BACKGROUND}

Although GaN is close to being a
polytypic material \cite{rev,fz}, we have not attempted a prediction of the
relative stability of ZB and wurtzite structures \cite{mra}. Such a prediction
would be somewhat secondary to our understanding of this system. The first
reason for this is that wurtzite-ZB energy differences
are known to be very small both experimentally and computationally in many
analogous systems \cite{nota}; second, wurtzite is already known to be the
stable equilibrium structure of GaN under normal experimental conditions;
finally, the
existence of an ``easy'' structural transitions induced by the substrate
symmetry in epitaxial growth has already been demonstrated
experimentally \cite{rev,expg}. Therefore, and also in view of the
existence of a synthetic cubic phase of GaN, we hold an analysis of the
properties of the cubic phase for more relevant at this stage.

In view of the real-space localization of the valence electrons of N and of
the $d$ shell of Ga, the use of an all-electron method, in which
electronic states are all treated on equal footing irrespective of their
localization properties, appears to be important. In particular, prior to
information to the contrary,
a proper treatment of the Ga $d$ shell is required.
To get an indication of the relevance of these states,
one can consider calculations for the free atoms.
The approximation of freezing any electronic states in the core, or
of eliminating them by a pseudization procedure, is only
plausible if they are clearly
more strongly bound than any of the relevant valence states in the free atom.
Further, even if the frozen-core or pseudoatom approximation appears to be
reasonable from inspection of the free atoms,
it may break down in a compound crystal in the presence of
resonances of the atomic levels  of different constituent atoms. An inspection
of the free-atom energy levels of Ga and N shows that a special situation
occurs in GaN (see Fig.\ref{uno}).
The upper valence bands will originate from the $sp$
states of Ga and the $2p$ electrons of N, the corresponding non-relativistic
full-core free-atom eigenvalues being $\epsilon_{4p}^{\rm Ga} = -0.20$  Ryd,
$\epsilon_{4s}^{\rm Ga} = -0.67$ Ryd,  $\epsilon_{2p}^{\rm N} = -0.53$ Ryd.
The lower $s$ valence band should stem from the $2s$ states of N, for which
$\epsilon_{2s}^{\rm N} = -1.35$ Ryd; most interestingly, though, the atomic
$d$-states of Ga are found almost in resonance with the N $2s$ states,
at $\epsilon_{3d}^{\rm Ga} = -1.47$ Ryd. From the atomic eigenvalues, one can
thus infer that the appearance
of $d$-bands must be expected near the lower valence band.
Some hybridization of the Ga $d$ states with all other valence
states is also to be expected. This suggests that the proper treatment of the
Ga $d$ states as bands is  important for understanding the properties of GaN.

\section{RESULTS}

The structural properties of GaN in the ZB structure from our
total-energy minimization are $a_0 = 8.44$ bohr ($\Omega$ = 150.3 bohr$^3$),
 $B = 1.98$ Mbar, and $E_{\rm coh}$ = 10.88 eV/cell (free atom
spin-polarization included).
 The reported experimental values of the lattice constant of
zincblende GaN range \cite{rev,expg}\, from 8.50 to 8.58 bohr, so that the
experiment-theory discrepancy is around 1\%, which is usual for LDA
calculations on semiconductors.
We assume the experimental cohesive energy per atom pair of the
wurtzite phase to be a reasonable estimate of the (as yet unknown)
experimental cohesive energy per cell of cubic GaN : this amounts to 9.0
eV/cell. The overestimation of cohesive energies is a known defect of
the LDA \cite{gd}.

To follow up the suggested special role of the cation 3$d$ states in GaN,
the band structures of three compounds are compared in Fig.\ref{due}:
those of ZnS, GaN, and GaAs.
The role of the metal $d$ states is strikingly evident in the GaN band
structure (central panel), which turns out to be
a borderline case between the typical
III-V compound GaAs (bottom) in which practically inert $d$ bands appear well
below the $s$-like bottom valence band, and the typical II-VI compound ZnS
(top), in which the $d$ bands lie high in the heteropolar gap.
In the case of ZnS, in spite of the relatively limited dispersion,
a large hybridization with
other states takes place (e.g., it has been shown \cite{cat}\,
for ZnO  that Zn 3$d$ character is as large as 30 to 70 \% in valence states
usually having mostly $sp$ character.)

It is apparent
that a proper treatment of the Ga $d$ shell is essential in determining the
band structure of GaN. As expected, the $d$ bands
are strongly hybridized with the bottom $s$-like valence band stemming
from N $2s$, which results in a large splitting away from the zone center
(as $sd$ mixing is symmetry-forbidden
at $\Gamma$). For example, at the L point the bottom valence band
(L$_1$ in Fig.\ref{due}) and the lower branch of the upper valence bands
(L$_1^v$ in Fig.\ref{due}) both belong to the L$_1$ representation of the
little group of {\bf k},
$C_{3v}$; the $d$ bands split into two basically inert L$_3^d$ doublets (one
of these originates from the non-bonding $\Gamma_{12}$, see Figures in Sec.V),
and an L$_1$ singlet which interacts strongly with the other L$_1$ bands
(see also the discussion in Sec.V).

In Fig.\ref{tre}, we
show a blow-up of the $d$ bands along the $\Lambda$ and $\Delta$
lines. Along $\Lambda$ the symmetry is $C_{\rm 3v}$ : the $\Gamma_{12}$ band
goes over to a $\Lambda_3$ state and remains two-fold degenerate, while the
$\Gamma_{15}$ splits into a $\Lambda_3$ doublet and a $\Lambda_1$ singlet,
which interacts with the bottom and top valence bands.
Along $\Delta$, the symmetry is $C_{\rm4v}$, so that $\Gamma_{15}$ splits into
a $\Delta_5$ doublet and a $\Delta_1$ singlet, and $\Gamma_{12}$ goes into
$\Delta_2$ and $\Delta_1$ singlets. The compatibility between the latter
and the singlet stemming from $\Gamma_{15}$ brings about an anticrossing
feature with antibonding and non-bonding character being exchanged between
the upper and lower level. The splitting of the hybridized band and
the center of mass of the ``inert'' $d$ bands is 2.03 eV at X and 1.56 eV at L.
It is of some interest to note that these features are an amplification of a
pattern present, although less pronounced, in the
$d$ bands of ZnS and ZnSe (we calculated these too, but we do not show
them here), in accordance with previous all-electron results \cite{zns}.

The (DFT-LDA) fundamental gap is found to be direct at $\Gamma$, $E_g$ = 2.0
eV. The experimental value of the fundamental gap is 3.55 eV in wurtzite
GaN, and reported values for cubic GaN range from 3.25 to 3.5 eV
\cite{rev,expg}.
An instant estimate of the self-energy correction to the gap \cite{gd,gss}\,
is provided by a recently proposed simple model \cite{fb}, which expresses it
as $\Delta \simeq 9/\varepsilon$ eV. Since the high-frequency dielectric
constant $\varepsilon$ is unknown as yet for the cubic phase, one can use the
wurtzite-phase $\varepsilon$ of 5.8, getting a corrected gap of 3.55 eV.
The calculated valence band width is 15.99 eV, which is substantially larger
 than that of the related compounds GaAs and GaP (13.0 and 13.1 eV
respectively). The main heteropolar gap ($s$ to $sp$ band) is 9.01 eV, while
the $d$-to-$sp$ gap is 5.39 eV (the heteropolar gap in GaAs is 3.44 eV). Apart
from $d$ hybridization, these large gaps are to be expected
in view of the ionicity of GaN.

Several pseudopotential calculations have been recently carried out for GaN,
but to our knowledge there were no all-electron calculation available on it
to date for the zincblende structure \cite{defp}. A proper
comparison of our results with previous theoretical investigations is
thus not straightforward, due to the problems encountered by pseudopotential
methods in
treating localized nodeless wavefunctions such as those of the N 2$p$
and Ga 3$d$ states. In the following we will compare our calculation with
results obtained by others for the zincblende structure only.

In the mixed-basis calculation of Ref.\cite{min}\, a pseudopotential
including non-linear core corrections (NLCC) \cite{nlcc}\, for the frozen
3$d$ shell
was used for Ga and the Bachelet-Hamann-Schl\"uter (BHS) potential \cite{bhs}
was employed for nitrogen. A cell volume of 134 bohr$^3$ (lattice constant
$a_0$ =
8.12 bohr, --3.7 \% with respect to our result), a bulk modulus of 2.4 MBar,
and a direct gap of 2.8 eV were reported. Assuming that the localized part of
the mixed basis can describe
reasonably well the nitrogen valence states despite the low plane-wave cutoff
of 14 Ryd, this severe underestimate of the cell volume could well be due to
the approximate treatment of the $d$-shell of Ga. This is consistent with past
experience with II-VI's \cite{wz}; we return to this question in the next
Section. We should mention though that the reported
cohesive energy is only 8.2 eV/cell; as has been argued in Ref.\cite{wz},
among the
effects of partially active $d$ shells there is an increase in lattice
constant {\it and} a reduction of the cohesive  energy  (see below), so we
would expect the pseudopotential cohesive energy to be larger than ours.
One may suppose then that this calculation may not be fully converged.

Recently, Palummo {\it et al.} \cite{pal} have studied GaN with plane
waves and pseudopotentials, treating the Ga $d$ shell as core states, and
linearizing the core-valence exchange-correlation functional (no NLCC).
Their results for zincblende GaN (a {\it tour de force} requiring a cutoff
of 120 Ryd) are somewhat puzzling. Using
the nitrogen pseudopotential by Gonze, Stumpf, and Scheffler (GSS) \cite{sgs}\,
they obtain $a_0 = 8.41$ bohr (--0.5\% from our result), $B=1.69$ MBar, and
$E_{\rm coh} = 10.25$ eV (--5.7 \% from ours),
whereas they find $a_0 = 8.15$ (--3.4\% from our result) and $B=2.4$ MBar when
using the nitrogen BHS potential, in close agreement with
Ref.\cite{min}.  The $s$ and $p$ channel of GSS potential and the
$s$, $p$, and $d$ channels of BHS potentials were used respectively,
the last channel listed being included in the local potential.
As the authors point out, this important test signals high sensitivity
of the results to the specific potential used for nitrogen, due to the
high ionicity of the system. The
fundamental gap reported is 2.70 eV, about 0.7 eV larger than ours
\cite{gpp}. This
discrepancy could also be attributed to the neglect of the Ga $d$ shell (see
Sec.IV).

A further comparison is possible with our calculated frequency of the TO
frozen phonon at the
zone center, which is $\Omega_{\rm TO}^{\rm ZB} = 600$ cm$^{-1}$. The
experimental frequency
is not known in cubic GaN, but the analogous mode in the wurtzite structure has
$\Omega_{\rm TO}^{\rm W} = 533$ cm$^{-1}$. Although the discrepancy is rather
large, part of the problem may lie in wurtzite-ZB differences. As an example,
the TO phonon frequency in cubic SiC is known experimentally
to be about 10 \% higher than in hexagonal SiC \cite{fz,sic}, so this
may account for most of the present deviation. On the other hand, the
pseudopotential result for the {\it wurtzite} TO frequency in
Ref.\cite{min} suffers from a considerably worse discrepancy
($\Omega_{\rm TO}^{\rm W} = 644$ cm$^{-1}$).

\section{EFFECTS OF {\it d} ELECTRONS}

The previous Section has demonstrated that the Ga $d$ states
play an important role in the band structure of GaN, due to their
strong hybridization with the N 2$s$ states near the bottom of the
valence band. Not yet clear is the importance of these active $d$ bands for the
cohesive properties and the charge density. We consider
these questions in this Section, coming to the conclusion that
GaN is quite similar to the II-VI compounds in these respects.

\subsection{Cohesive properties}

The question to be answered is whether a reasonable description
of the bonding in GaN can be obtained with an approach which only
treats the $sp$ electrons of Ga and N explicitly as band states.
In the pseudopotential
framework, for example, this could be done by simply treating the
cation $d$ states as core states and linearizing core-valence
exchange-correlation. However, this approach has been
shown to fail manifestly for the II-VI compounds, giving lattice
constants which are typically 10 to 15 \% too small \cite{wz,eng}. In this
respect, improved   pseudopotential techniques  adopt the so-called non-linear
core corrections (NLCC) \cite{nlcc}.
Hereby the free-atom core density is added to the valence density
when the exchange-correlation energy is calculated. This improves
things considerably for the II-VI's, since the lattice constant is now
only 3 to 4.5\% below the experimental value \cite{eng,kle}.

More generally, the core states can affect the bonding in three distinct
ways.
%
{\it First}, if the core
is frozen to its free-atom shape, an (usually small) error is made
which, according to the variational principle, must increase the total energy,
and hence decrease the cohesive energy. This effect disappears when the
lattice constant approaches infinity, and it increases monotonically for
decreasing lattice constant as long as the core states remain very localized
on the typical length scale of the lattice : in the latter regime this effect
leads then to an increase of the lattice constant.
{\it Second}, if the core overlaps appreciably with the on-site valence
states, the effect of the exchange-correlation non-linearity sets in,
resulting in an interatomic repulsion.
{\it Third}, if the core states are large enough to overlap to some extent
(or to interact with neighboring-site filled valence states, as is the case
for the Ga $d$s and N $s$ in GaN), the closed-shell
repulsion becomes noticeable.
This is a consequence of the increase in the
kinetic energy when the cores on neighboring sites are made orthogonal.
If this contribution is neglected, the lattice constant comes
out too small and the cohesive energy too large. The magnitude of this effect
is governed not only by the overlap of the core wavefunctions, but also by
the amount of resonance between the on-site core eigenvalues for the
neighboring sites.
This means that the core states can only ``see'' each other if they are
in the same energy range. For a system with neighboring atoms of different
types, it is therefore conceivable that there is a reduced closed-shell
repulsion if core eigenstates
do not resonate. Conversely, for the case of GaN we expect a strong such
repulsion  since, as we have seen, there is a very strong resonance between
the Ga 3$d$ and N 2$s$ states.

To distinguish between the various core contributions to the bonding,
we have done additional calculations as follows. In each iteration,
the core density is obtained by overlapping the frozen free-atom
cores, and the sum of the free-atom core kinetic energies is taken
into the total energy. The overlapped core can extend into the
interstitial region and into neighboring atomic spheres. In cases
where the core states are localized enough not to overlap,
this ``frozen--overlapped-core'' approximation (FOCA) is equivalent to the
usual frozen-core approximation. For GaN, however, the Ga 3$d$ states are so
large compared to the interatomic distance that
a straight frozen-core calculation runs into problems. Of the three
core contributions discussed above, the FOCA procedure includes only the
non-linear exchange-correlation contribution; it does not include
the core relaxation and closed-shell repulsion effects. It should
therefore be closely similar to the pseudopotential-plus-NLCC
approach, and should shed light on the applicability of the latter to GaN
and related systems.

Table I shows the cohesive properties of GaN, ZnS, GaAs, and Si
calculated by the all-electron FP-LMTO method, both with the full treatment
and using the FOCA, and compares them to NLCC-pseudopotential results where
these are available. For Si, the core is very small so that the only neglected
term in the FOCA is the core relaxation. In agreement with the discussion
above,
this leads
to a slightly larger lattice constant and a smaller cohesive energy.
A similar behavior is seen for GaAs, for which the Ga $d$ states are
energetically well below the As $s$ states and, in addition, the lattice
constant is rather large. The decoupling of the $d$ states is not complete,
though, and the relatively small effects of core relaxation and closed-shell
repulsion are in competition.

For ZnS as well as GaN, on the other hand, the relevant effect neglected in
the FOCA is the closed-shell repulsion with the associated underestimate of
the lattice constant and overestimate of the cohesive energy. Although
acceptable values are obtained, the lattice constant is still too small by
some percent. This is in fact similar to the deviations found when the
pseudopotential technique with NLCC is applied, as can be seen from Table I.
We note though that the NLCC-PP errors tend to be somewhat larger. This may be
attributed in part to the additional approximations needed in the
implementation
of the NLCC (e.g. the pseudization of the core charge inside some inner core
radius).

On the basis of the above considerations
we conclude that in the case of GaN no pseudopotential calculations can be
considered reliable if performed without the $d$ shell in the valence or
(as a minimal requirement) use of the NLCC. This is also the case for II-VI
materials, as shown above and in earlier work \cite{wz,eng,kle}. Satisfactory
results obtained in this context with
the direct, uncorrected pseudopotential procedure should thus be regarded as
incidental \cite{pal,kun}, and possibly due to accidental error cancellation.

We briefly consider the influence of Ga $d$-shell freezing
on the band structure of GaN, in particular on the direct gap, which is
2.00 eV in the full calculation.
Although the absence of repulsion between $d$ and top valence $p$ states at
zone center causes an increase of the gap (see the detailed discussions in
Ref.\cite{zns,wz}),
a quantitative estimate of gap differences is difficult, due to
volume dependences; the high bulk modulus and
deformation potentials of the relevant states  cause effects similar to those
observed for other semiconductors \cite{me2}. A gap of 2.8 eV is
reported in Ref.\cite{min}, and one of 2.7 eV in Ref.\cite{pal}.
In our frozen-core calculation we find a
gap of 2.66 eV at the frozen-core theoretical lattice constant (8.30 bohr).
When calculated at the theoretical lattice constant of the full calculation
(8.44 bohr), the frozen-core gap is 2.20 eV, 10 \% larger than the full
gap at the same volume. We report in passing our estimate for the deformation
potential of the lowest conduction state, 5 meV/kBar, in good agreement
with the experimental value of 4.7 meV/kBar \cite{defp}.

As a summary of this Subsection, the core states can influence the
bonding properties in three ways, namely through
non-linear core-valence exchange-correlation effects, core relaxation, and
closed-shell repulsion. To obtain a resonable description of bonding in
II-VI and III-V semiconductors with $d$ states in the valence range, it is
imperative to include non-linear core corrections in a pseudopotential
treatment. The core relaxation has a minor influence and can normally
be neglected. The closed-shell repulsion is rather significant
in those cases for which the cation $d$ states are in the same
energy range as the anion $sp$ states. In this sense, GaN turns
out to be very similar to the II-VI compounds. For all of these materials,
a full treatment of the $d$ electrons as band states is needed in
order to reduce the remaining 3-4 \% underestimate of the lattice constant
to the LDA-standard 0.5-1\%.

\subsection{Wavefunctions and densities}

The similarity of GaN and the prototypical II-VI ZnS has been observed in
the band structure and cohesive properties. Such effects are due to the
cation $d$ shell being not inert in these systems.
To render this visually, we present a selection of partial charge densities for
GaN. This is similar in spirit to the work of Wei and Zunger \cite{wz}, who
have investigated in some detail the effect of cation $d$ states on (among
others) the charge densities of selected states and on the total densities in
several II-VI compounds. We will be considering charge densities at specific
k-points (the modulus squared of the wave function $\psi_i$ ({\bf k, r}) at
{\bf k}) as well as k-integrated band contributions to the density. All charge
densities are plotted in a (110) plane. For convenience, superscripts
$d$ and $v$ are used to label states originating from $d$ and top valence
bands respectively, whereas states with no superscripts are from the bottom
valence band (see Fig.\ref{due} for a guide to the relevant states). Density
plots have lowest contour and contour spacing of 0.5 and 1 electron/cell,
respectively.

In the zincblende structure, the mixing of $p$ and $d$ states is permitted
\cite{wz}\, even at the $\Gamma$ point, which has $T_d$ as group of {\bf k}
(contrary to the situation in the diamond structure, where it has the full
 $O_h$ symmetry). The upper valence bands at $\Gamma$ will no longer have
pure $p$ character, $d$ character being admixed into them and, conversely, the
$d$-bands will contain $p$ components. The mixing of $s$ and $d$ states is
not allowed at $\Gamma$, but at a general point of lower symmetry, states with
all symmetries of the rotation group will in general be allowed to interact.
Therefore, charge density and density of states (which are obtained by
averaging
over the Brillouin zone), and the full band structure
will be only partially reminiscent of the specific symmetry of the original
atomic states. This is the case, $e.g.$, for the
$\Gamma_{15}^d$
triplet of $d$ origin which is only $p-$admixed at the zone center, but
gives rise to a strongly $s-$ and $p-$admixed singlet away from the zone
center, and for the $\Gamma_1$ $s$-like singlet, which is similarly admixed
with $pd$ character away from $\Gamma$. Already at the zone center, the
ordering of the $d-$like $\Gamma_{15}$
triplet and $\Gamma_{12}$ doublet (which remains practically inert throughout
the zone, resulting in a non-bonding state)
is reversed with respect to what is expected \cite{fig}\, for a
tetrahedral environment, due to  $pd$ repulsion with the top valence band
(qualitative schemes of this interaction are given in Ref.\cite{wz}).

In Fig.\ref{quattro} we plot the squared wavefunctions at the zone center for
the states $\Gamma_{15}^d, \Gamma_{12}^d$ and $\Gamma_{15}^v$ in GaN.
The former two states originate
from the tetrahedral splitting of atomic $d$ states (the 5-fold degenerate $d$
multiplet is split in a $\Gamma_{15}$ triplet and a $\Gamma_{12}$ doublet),
the latter is the bonding $p-$like valence state. The $\Gamma_{12}$ state
remains very similar to a $d_{z^2-r^2}$-like atomic orbital and forms a
non-bonding, inert state. In the other states, the prevailingly ionic character
of the bonding is visible; the admixture of $d_{xz}$-like and $p$ character
results in
a weak bonding contribution to  $\Gamma_{15}^d$ and an antibonding one to
$\Gamma_{15}^v$.

While at zone center only $pd$ mixing is allowed, this restriction
does not apply to lower-symmetry points. We consider as representative the
squared wavefunctions at the L point. At that k-point the bottom valence band
is of L$_1$ symmetry; the $\Gamma_{15}^d$ triplet splits further into an
L$_3^d$ doublet, analogous to the $(\Gamma_{12})$-L$_3^d$, and an L$_1^d$
singlet. Figure \ref{cinque}
shows the symmetrized squared wavefunction at L (the average over the eight
equivalent L-points in the Brillouin zone), for
the states L$_1$, L$_1^d$, L$_1^v$, L$_3^v$ (top to bottom). We
notice that the bottom valence L$_1$ has a bonding $d$ admixing, and L$_1^d$
and L$_3^v$ are instead antibonding.

Next, in Fig.\ref{sei} we plot the k-integrated partial charge densities
$n_i ({\bf r})$ for
the bottom, $d$, and top valence bands. The first two are $sd$-like and
have some bonding  contribution from the $d$-like states, the latter
is ionic $p$-like and has an antibonding $d$-admixing, which is visible as
a depletion along
the bond direction. As expected the Brillouin zone summation has brought about
a mixing of all angular momenta (in
particular $sd$ in the two lower bands). The top valence states are clearly
dominated by ionic charge transfer.

Finally, we compare GaN to ZnS and GaAs in Fig.\ref{sette}.
Shown for the three materials (top to bottom : ZnS, GaN, GaAs) are the states
$\Gamma_{15}^v$ (right) and the k-integrated densities of the bottom
valence band (center) and of the top valence band complex (left).
The similarity of GaN and ZnS is remarkable.
The effect of $d$ admixing is more visible in the ZnS
top valence charge, as the $d$-bands are close to and significantly hybridzed
with the top valence bands \cite{cat}.
The bonding contribution of the $d$ electrons to the ionic
bottom valence charge is instead stronger in GaN, as here the hybridization
of $s$ and $d$ bands is larger.
The top valence band, zone-center state $\Gamma_{15}^v$ (right) shows a
conspicuous
antibonding contribution of the $d$ electrons, of a similar kind in ZnS and
GaN. Even in GaAs, though, a slight $d$-admixing is visible in this state
(analogous to that observed recently in InP, InAs, and InSb \cite{mas}),
and its bottom valence charge also shows some $d$ component.
We should mention that ionicity is only partially
responsible for the differences between GaAs and GaN, or the similarities of
ZnS and GaN, as the ionicity of GaN
is 0.43 as compared to 0.33 of GaAs and 0.70 of ZnS \cite{defp}.
The features of the partial densities should also
be directly compared with the results of Ref.\cite{wz}\, for II-VI compounds.

We conclude that in contrast to most other III-V compounds (represented
here by GaAs), GaN does exhibit features of the charge densities and band
structure which resemble those observed \cite{wz} for II-VI's (represented here
by ZnS). This is due to the $d$ shell of Ga being not inert in this compound,
as a consequence of the energy resonance of N $s$ and Ga $d$ levels.

\subsection{Density of states}

Although mixing of all angular momenta takes place due to symmetry lowering in
the crystal environment, the content of a specific angular momentum character
of a
crystal state may be recognized in several ways. As seen above, inspection of
partial charge densities is useful in this respect. Another technique allowing
a monitoring of the site and angular momentum dependence of the charge
distribution is the Mullikan decomposition \cite{mull} of the
total density of states,  which is natural to the LCAO-like basis used
here. The total DOS can be decomposed into site and angular
momentum contributions, and if desired, into bonding and antibonding overlaps.

The LMTO basis set naturally gives a decomposition of the calculated
wavefunctions as a sum over sites and angular momenta as
$\psi_{\bf k} = \sum_{\nu,L} \psi_{\nu,L}^{[{\bf k}]}$\, , where
$\nu$ is a site index and $L$ a composite angular
momentum index. We then take the decomposition of the norm as weight to add
into the partial densities of states as
\begin{equation}
1 = \langle\psi_{\bf k}|\psi_{\bf k}\rangle = \sum_{\nu,L}
{\tt Re} \left[\langle\psi_{\bf k} |\psi_{\nu,L}^{[{\bf k}]}\rangle\right] .
\end{equation}
The advantage of this approach over the common technique of projecting onto
angular momenta inside the spheres is obvious : first, the decomposition is
essentially independent of all sphere radii, and second, no DOS contribution
is associated with the empty spheres or the interstitial region.

The atom-projected DOS for $s$, $p$, and $d$ angular momenta are given in
Fig.\ref{otto}. A gaussian broadening of 0.02 Ryd and a mesh of 344
special points in
the irreducible wedge of the Brillouin zone are used. The $s$ and $d$ bands
have contributions from both Ga and N sites, but this is a consequence of their
having exchanged
characters to a considerable extent. Most $s$ character comes from N, whereas
most of the $d$ character resides on the Ga site, and is barely sizable
on N over the whole valence band span (the total charge of $d$ character
contributed by  the N site is about 0.08 electrons). The N $p$ states dominate
the top valence band, although of course $spd$ Ga contributions are visible.
The $s$ DOS shows a predominant contribution of N $2s$ to both  of the lower
bands, and almost none in the upper bands, where some Ga $s$ character is
present instead.

\section{SUMMARY}

We have calculated the properties of $\beta$-GaN using the {\it ab initio}
all-electron full-potential LMTO method. Agreement is found with existing
experimental data, and predictions are made for other quantities.
The explicit consideration of the $d$ electrons of Ga has been shown
to be of importance in determining the properties of the material.
Due the resonance of N $s$ and Ga $d$ states, the band structure, DOS, and
charge density of GaN are quite peculiar for a III-V compound, as
significant 3$d$
character is present in all the valence bands. Several features of GaN of
the band structure, charge density, and density of states do
actually resemble those of II-VI materials.

A frozen--overlapped-core procedure was used to identify the way in which the
core and $d$ states influence the cohesive properties of GaN as compared to
some other relevant materials. We deduce that, although
the non-linear exchange-correlation correction is the most important effect in
GaN and ZnS, the closed-shell repulsion also plays a significant role; neglect
of this
effect leads to a lattice constant which is too small by some 3\% in both
systems. This sets an upper limit on the accuracy which can be expected
of methods which do not treat the $d$ states esplicitly. Therefore, even
pseudopotentials including non-linear core corrections, although going a long
way in the right direction, are generally not sufficient for a description
of GaN, as well as of ZnS and similar system, at the standard LDA level
of accuracy for semiconductors.


\newpage

\figure{Non relativistic free-atom LDA eigenvalues for
some elements involved in the formation of III-V and II-VI compounds.
\label{uno}}

\figure{FP-LMTO band structure of GaN (center),
compared to those of ZnS (top), and GaAs  (bottom). State labels for
the central panel apply straightforwardly to
the others. The energy zero is the valence band
top. The bands are calculated at the theoretical lattice constants.
\label{due}}

\figure{Detail of the $d$ bands of GaN along L-$\Gamma$-X.
Degeneracies are indicated in parentheses. \label{tre}}

\figure{Squared wavefunctions of the states $\Gamma_{15}^v$ (top),
$\Gamma_{12}^d$ (center), $\Gamma_{15}^d$ (bottom) at the zone center in GaN.
\label{quattro}}

\figure{Symmetrized squared wavefunctions of
the states L$_3^v$, L$_1^v$, L$_1^d$, and L$_1$ (top to
bottom) at the L point in GaN. \label{cinque}}

\figure{Partial k-integrated charge densities in GaN for the
top valence bands (top), $d$ bands (center) and bottom valence band (bottom).
\label{sei}}

\figure{ Comparison of ZnS (top row), GaN (center) and GaAs (bottom).
Right column : state $\Gamma_{15}^v$; central column : bottom valence band
partial density; left column : top valence band partial density.
\label{sette}}

\figure{ Mullikan angular momentum decomposition of
atom-projected densities of states.
 From top to bottom : Ga $s$, $p$ and $d$
states, and N $s$, $p$ and $d$ states. \label{otto}}

\newpage
\widetext

%
\begin{table}
\begin{tabular}{cc|rrrrrrr}
\multicolumn{2}{c}{ } &
\multicolumn{2}{c}{ all-electron$^a$ } &
\multicolumn{2}{c}{ FOCA$^b$ } &
\multicolumn{2}{c}{ NLCC-PP$^c$ } &
\multicolumn{1}{c}{ Exp.$^d$ }\\
\tableline
GaN  & $a_0$ &   8.44 & (--1.2) &  8.30 & (--2.8) &  8.12 & (--4.7) &  8.54 \\
     & $B$     &  1.98 &        &  2.00 &        &  2.40 &        &  --   \\
     & $\delta{E}_{\rm coh}$ &  &        &  0.33 &        &   &   &   \\
\tableline
ZnS  & $a_0$ & 10.13 & (--0.8) &  9.92 & (--2.9) &  9.80 & (--4.0) & 10.21 \\
     & $B$     &  0.75 &        & 0.83 &        &  1.06 &        &  0.77 \\
     & $\delta{E}_{\rm coh}$ &  &        &  0.34 &        &   &   &   \\
\tableline
GaAs & $a_0$ & 10.62 & (--0.6) & 10.71 & (+0.2) &  10.50 & (--1.7) & 10.68 \\
     & $B$     &  0.75 &        &  0.65 &        &  0.77 & &  0.77 \\
     & $\delta{E}_{\rm coh}$ &  &        &  0.08 &        &   &   &   \\
\tableline
Si & $a_0$ & 10.22 & (--0.4) & 10.24 & (--0.2) &  & & 10.26 \\
     & $B$     &  0.96 &        &  0.96 &        &  & &  0.99 \\
     & $\delta{E}_{\rm coh}$ &  &        &  --0.41 &        &   &   &   \\
\end{tabular}
\caption{Calculated structural properties for GaN, ZnS, GaAs,  and Si. The
 lattice constant $a_0$ is in atomic units, $B$ is the bulk modulus in Mbar,
$\delta E_{\rm coh}$ = $E^{\rm frozen}_{\rm coh} - E^{\rm full}_{\rm coh}$
is the difference of cohesive energies (eV/cell) in the frozen-core and full
all-electron calculations, respectively.
Numbers in parentheses indicate deviations from experimental data.}
\tablenotes{
{\it a} : this work, full FP-LMTO \\
{\it b} : this work, frozen--overlapped-core FP-LMTO \\
{\it c} : pseudopotentials with non-linear core corrections
(Ref.\cite{min}\, for GaN, Ref.\cite{eng}, for
ZnS, Ref.\cite{mp}\, for GaAs) \\
{\it d} : data from Land\"olt-Bornstein : Numerical Data and Functional
Relationships in Science and Technology, New Series Group III,
K.- H. Hellwege and O. Madelung eds., (Springer, New York, 1982),
Vol.17a and 22a and references therein. For cubic GaN an average of
the reported experimental data \cite{rev,expg}\, is used.}
\end{table}

\newpage

\begin{references}
\bibitem{rev}
R. F. Davis, Proceedings IEEE {\bf 79}, 702 (1991).
\bibitem{expg}
M. J. Paisley, Z. Sitar, J. B. Posthill, and R.F. Davis, J. Vac. Sci. Technol.
{\bf A\,7}, 701 (1989);
R. C. Powell {\it et al.},
Material Research Society Symposia Proceedings, vol.{\bf 162}, J.T. Glass, R.F.
Messier and N. Fujimori eds., pag.525 (Material Research Society,
Pittsburgh, PA, 1990);
M. Mizuta, S. Fujieda, Y. Matsumoto, and T. Kavamura, Jap. J. Appl. Phys.
{\bf 25}, L945 (1986);
G. Martin, S. Strite, J. Thornton, and H. Morkoc, Appl. Phys. Lett. {\bf 58},
2375 (1991).
\bibitem{gd}
R. Dreizler and E. K. U. Gross, {\it Density functional theory}, (Springer
Berlin, 1990).
\bibitem{fp}
M. Methfessel, Phys. Rev. {\bf B 38}, 1537  (1988); M. Methfessel, C. O.
Rodriguez, and O. K. Andersen, Phys. Rev. {\bf B 40}, 2009 (1989). See also
M. Methfessel and M. Scheffler, Physica {\bf B 172}, 175 (1991),
for an application to semiconductor interfaces.
\bibitem{vwn}
S. H. Wosko, L. Wilk, and M. Nusair, Can. J. Phys. {\bf 58}, 1200 (1980).
\bibitem{ca}
D. M. Ceperley and B. J. Alder, Phys. Rev. Lett. {\bf 45}, 566 (1980).
\bibitem{mpa}
H. J. Monkhorst and J. D. Pack, Phys. Rev. {\bf B  13}, 5188 (1976).
\bibitem{asa}
O. K. Andersen, O. Jepsen, and D. Gl\"otzel, in {\it Highlights of
Condensed Matter Theory}, F. Bassani, F. Fumi, and M. P. Tosi eds.,
(North-Holland, Amsterdam 1985).
\bibitem{fz}
See C.Y. Yeh, Z. W. Lu, S. Froyen, and Alex Zunger, Phys. Rev. {\bf B
45}, 12130 (1992), and references therein.
\bibitem{mra}
The feasibility of FP-LMTO calculations in strained and complicated geometries
for semiconductors has been previously demonstrated \cite{fp}.
 \bibitem{nota}
Typical calculated
total energy differences between wurtzite and zincblende are fractions
of mRy per atom, and are quite sensitive to technicalities as k-point summation
(for example a too coarse k-points mesh tends to spuriously
stabilize the zincblende structure \cite{fz})
\bibitem{cat}
A. Catellani, A. Baldereschi, and M. Posternak, unpublished.
\bibitem{zns}
The features of the bands of ZnS and GaAs are in quite good agreement with
previous calculations (see, e.g., the LMTO calculation by
G. B. Bachelet and N. E. Christensen, Phys. Rev. {\bf B 31}, 879 (1985),
for
GaAs, and the FLAPW results of A. Continenza, S. Massidda, and
A. J. Freeman, Phys. Rev. {\bf B 38}, 12996 (1988), for ZnSe).
LAPW and pseudopotential calculations (the latter including the $d$ electrons
in the valence) have been performed for ZnS by J. L. Martins, N. J. Troullier,
and S.-H. Wei, Phys. Rev. {\bf B 43}, 2213  (1991).
\bibitem{gss}
See, e.g., R. W. Godby, L. J. Sham, and M. Schl\"uter,
Phys. Rev. {\bf B 37}, 10159 (1988).
\bibitem{fb}
V. Fiorentini and A. Baldereschi, J. Phys. Condens. Matter {\bf 4}, 5967
(1992);
Solid State Commun., submitted.
\bibitem{defp}
For the wurtzite phase, an experimental and theoretical investigation
has been carried out by
P.Perlin, I. Gorczyca, N. E. Christensen, I. Grzegory, H. Teisseyre, and
T. Suski, Phys. Rev {\bf B 45}, 13307 (1992), using the LMTO-ASA method, and
FLAPW calculations have been performed by A. Catellani, A. Baldereschi, and
M. Posternak (unpublished, private communication).
\bibitem{min}
B. J. Min, C. T. Chang, and K. M. Ho,  Phys. Rev {\bf B 45}, 1159 (1992).
\bibitem{nlcc}
S. Froyen, S. G. Louie, and M. L. Cohen, Phys. Rev. {\bf B  26}, 1738 (1982).
\bibitem{bhs}
G. B. Bachelet, D. Hamann, and M. Schl\"uter, Phys. Rev. {\bf B 26}, 4199
(1982).
\bibitem{wz}
S.-H. Wei and A. Zunger, Phys. Rev. {\bf B 37}, 8958 (1988).
\bibitem{pal}
M. Palummo, C. M. Bertoni, L. Reining, and F. Finocchi, to appear in Physica
{\bf B}
\bibitem{sgs}
X. Gonze, R. Stumpf, and M. Scheffler, Phys. Rev. {\bf B 44}, 8503 (1991).
\bibitem{gpp}
Of course an accurate estimate of the LDA gap is essential to the calculation
of
quasiparticle energies.
\bibitem{sic}
L. Patrick, D. R. Hamilton, and W. J. Choycke, Phys. Rev. {\bf 143}, 526
(1966).
\bibitem{eng}
G. E. Engel and R. J. Needs, Phys. Rev. {\bf B 41}, 7876 (1990).
\bibitem{kle}
A. Kley, J. Neugebauer, and M. Scheffler, unpublished.
\bibitem{mp}
M. Peressi, private communication.
\bibitem{kun}
P. E. Van Camp, V. E. Van Doren and J. T. Devreese, Solid State Commun.
{\bf 81}, 23 (1992); A. Mu\~noz and K. Kunc, Phys. Rev {\bf B 44}, 10372
(1991).
\bibitem{me2}
V. Fiorentini, Phys. Rev. {\bf B 46}, 2086 (1992); Solid State
Commun. {\bf 83}, 871 (1992).
\bibitem{fig}
B. Figgis, {\it Introduction to ligand fields theory}, (Interscience, New York,
1966).
\bibitem{mas}
S. Massidda, A. Continenza, A. J. Freeman, T. M. De Pascale, F. Meloni, and
M. Serra, Phys. Rev. {\bf B 41}, 12079 (1990).
\bibitem{mull}
R. S. Mulliken, J. Chem. Phys. {\bf 23}, 1833 (1955);
T. Hughbanks and R. Hoffmann, J. Am. Chem. Soc. {\bf 105}, 3528 (1983).
\end{references}
\end{document}